\documentclass[twocolumn]{svjour3}

\usepackage[T1]{fontenc}
\usepackage[utf8]{inputenc}
\usepackage{ae,aecompl}

\usepackage{graphicx}	% Including figure files
\usepackage{amsmath}	% Advanced maths commands
\usepackage{amssymb}	% Extra maths symbols
\usepackage[square]{natbib}
\usepackage{hyperref}
\hypersetup{
    colorlinks=true,
    linkcolor=blue,
    filecolor=magenta,      
    urlcolor=cyan,
}

\journalname{Experimental Astronomy}
\begin{document}\sloppy

\title{An upper limit calculator (UL-CALC) for undetected extended sources with radio interferometers: radio halo upper limits}

\author{Lijo T. George        \and
        Ruta Kale \and %etc.
        Yogesh Wadadekar
}

\authorrunning{George et. al.}
\titlerunning{UL-CALC}

\institute{L. T. George \at
              National Centre for Radio Astrophysics - Tata Institute of Fundamental Research \\
              Ganeshkhind, Pune - 411007 \\
              \email{lijo7george@gmail.com}           %  \\
           \and
           R. Kale \at
              NCRA-TIFR \\
              Ganeshkhind, Pune - 411007
        \and
            Y. Wadadekar \at 
                NCRA-TIFR \\
                Ganeshkhind, Pune - 411007
}

\date{Received:  / Accepted: }

\maketitle

\begin{abstract}
Radio halos are diffuse, extended sources of radio emission detected primarily
in massive, merging galaxy clusters.
In smaller and/or relaxed clusters, where no halos are detected, one can instead place
upper limits to a possible radio emission. Detections and upper limits are both crucial 
to constrain theoretical models for the generation of radio halos. 
The upper limits are model dependent for radio interferometers and thus the process of 
obtaining these is tedious to perform manually. In this paper, we present a Python based tool 
to automate this process of estimating the upper limits. 
The tool allows users to create radio halos with defined parameters like physical size, 
redshift and brightness model. A family of radio halo models with a range of flux densities, 
decided based on the rms noise of the image, is then injected into the parent visibility file and imaged.
The halo injected image and the original image are then compared to check for the radio halo detection 
using a threshold on the detected excess flux density. 
Injections separated by finer differences in the flux densities are carried out 
once the coarse range where the upper limit is likely to be located has been identified. 
The code recommends an upper limit and provides a range of images for manual inspection. 
The user may then decide on the upper limit. We discuss the advantages and limitations of this tool. 
A wider usage of this tool in the context of the ongoing and upcoming all sky surveys with the LOFAR and SKA 
is proposed with the aim of constraining the physics of radio halo formation. 
The tool is publicly available at \url{https://github.com/lijotgeorge/UL-CALC}.


\end{abstract}

%%%%%%%%%%%%%%%%%%%%%%%%%%%%%%%%%%%%%%%%%%%%%%%%%%

%%%%%%%%%%%%%%%%% BODY OF PAPER %%%%%%%%%%%%%%%%%%

\section{Introduction}

The radio interferometric technique used in producing images of the sky in radio bands 
uses the Fourier transform relation between the visibilities 
and the intensity distribution in the sky \cite[Chapter 15]{tms17}.
 
The image produced by inverse Fourier transforming the visibilities is 
limited to the angular scales ($\theta$) which are sampled by the baselines of the interferometer.
For a range of baseline lengths, from $B_{min}$ to $B_{max}$, the smallest angular scale is
$\theta_{min} = \lambda/B_{max}$ and the largest scale is $\theta_{min} = \lambda/B_{min}$, 
where $\lambda$ is the wavelength. It has been shown that radio interferometers 
recover a fraction of the total flux density at a given angular scale even 
when that scale is sampled. Good overall uv-coverage with full synthesis imaging
at that angular scale is
needed for a near complete flux recovery (\cite{deo17}). 
The effect of insufficient sampling in the uv-plane is especially evident 
in the study of extended radio sources - for example, radio halos in 
galaxy clusters (\cite{kal15}) and supernova remnants (\cite{nayana17}). 
In the case of non-detection of such sources, the upper limit is dependent on the uv-coverage of
that particular observation and needs to be obtained using models of the brightness distribution. 

The method of injecting models of the expected extended source and 
quantifying the recovery has been extensively used to obtain the upper limits for the detections of radio halos. 
Radio halos are diffuse radio sources of synchrotron origin that are associated, 
not with individual galaxies, but with the intra-cluster medium (see \cite{wee19} for a review) in galaxy clusters. They have linear extents of 1 - 2 Mpc that correspond to angular extents of a few to tens of arcminutes in clusters at redshifts $<0.4$. They are also rare and thus the rates of non-detection in samples of clusters can exceed $70\%$. The non-detection of a radio halo is highly dependent on the %uv-coverage 
short spacings in the data in addition to the rms sensitivity. The method of obtaining model dependent upper limits was first used in the GMRT Radio Halo Survey (\cite{ven07,ven08}) and then in the Extended GMRT Radio Halo Survey (\cite{kal13,kal15}). The steps of 
building a model radio halo, injecting it in the visibilities and imaging those to check the recovery were carried out either manually or in a semi-automated fashion making it a slow and tedious process.

In this work, we present a general tool that we call the ``UL-CALC'' to obtain upper limits on extended sources of a model of
the user's choice. Application of this tool is illustrated in the context of finding upper limits  for radio halos in galaxy clusters, and discussed in the context of other areas such as mini-halos,
radio relics, supernova remnants and radio galaxies. The paper is organised as follows: 
the models implemented in literature for radio halos are described in Section~\ref{models}. 
In Section~\ref{haloc} the halo upper limit calculator is presented and an illustration of its
usage is provided in Section~\ref{usage}. We discuss the advantages and caveats of the calculator and its applicability to large surveys in Section.~\ref{disc}. Conclusions are presented in
Section.~\ref{conc}. 

\begin{figure*}
    \centering
    \includegraphics[width=\textwidth]{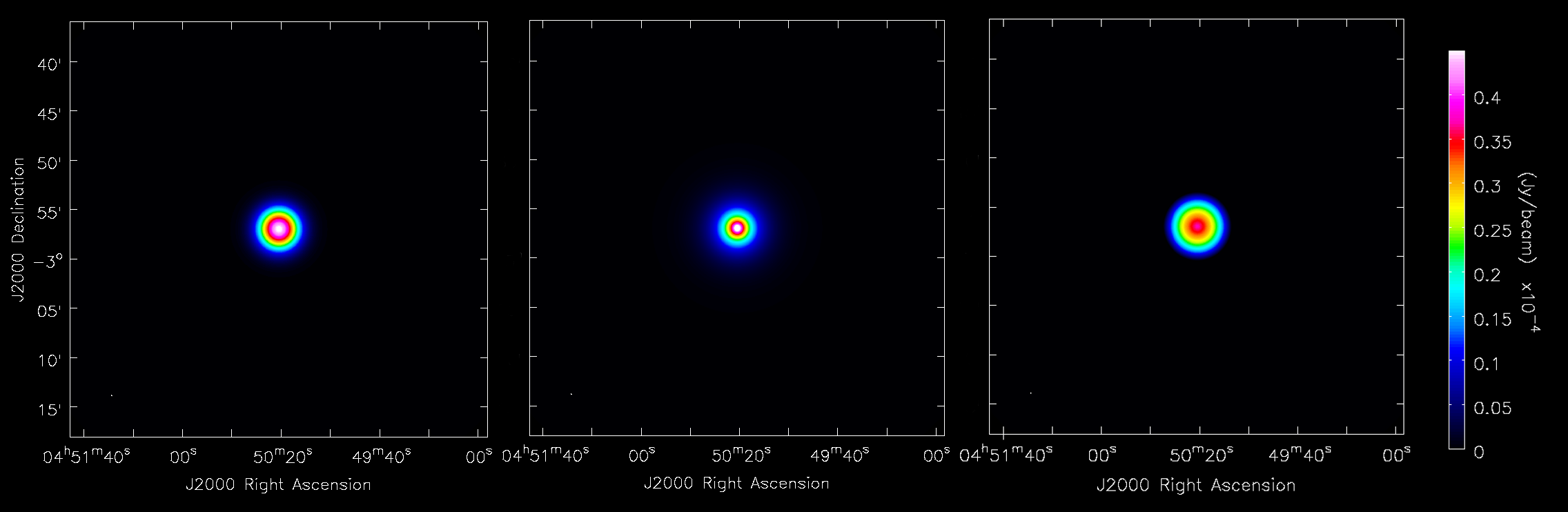}
    \caption{The various halo models available in the upper limit calculator. From left to right, they are the Gaussian, Exponential and Polynomial models.
    A colorbar common to all three images is shown to the right as well.
    All the halos have a physical size of 1000~kpc, total intensity of 1~Jy 
    and are placed at a redshift of 0.1.}
    \label{fig:models}
\end{figure*}

\begin{figure}
    \centering
    \includegraphics[width=\columnwidth]{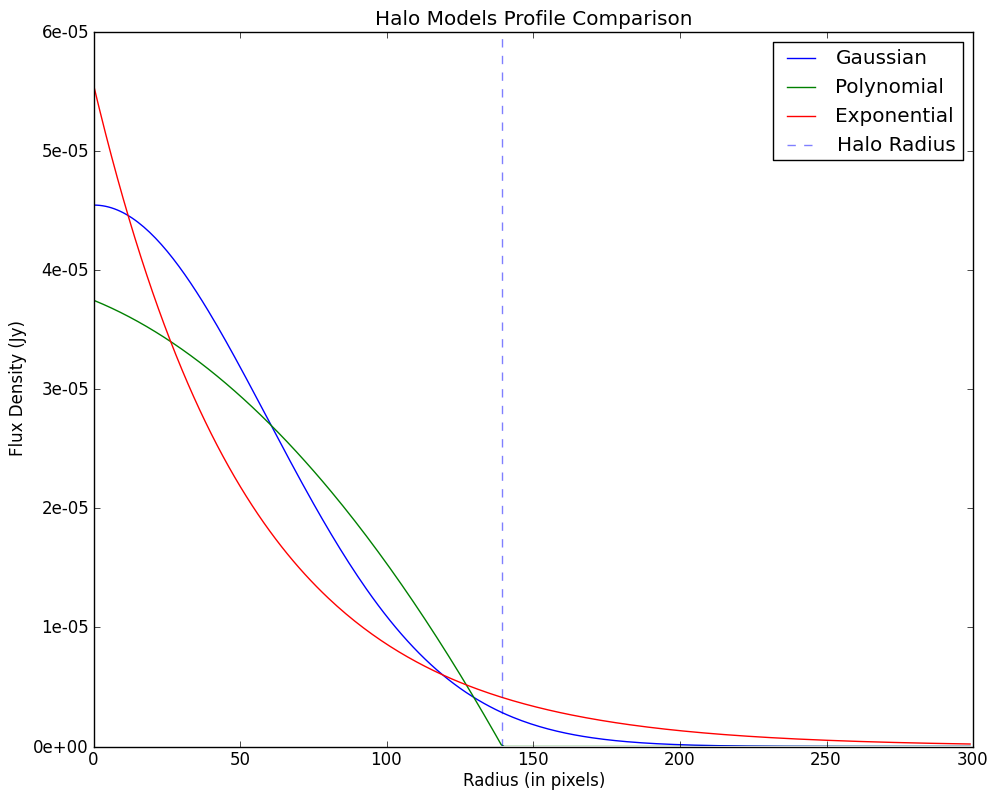}
    \caption{Radial profile comparison of the various halo models as depicted in Figure~\ref{fig:models}. The dashed line corresponds to the radius of the halo calculated
    in pixels ($139.33$) assuming a physical size of 1 Mpc and a redshift of 0.1 and each pixel corresponds to $1.5"$}
    \label{fig:model_profile}
\end{figure}

\section{Models for radio halos}\label{models}
    Radio halos are of synchrotron origin and are thus manifestations of the cosmic ray electrons
    ($\sim$GeV) and magnetic fields ($\sim \mu$G) pervading the ICM (Intra Cluster Medium). These sources are known to occur in about $22\%$ of
    massive ($>5\times10^{14}$ M$_\odot$) and merging galaxy clusters (\cite{kal15}). 
    Although there is a large variety in the details of the morphology of radio halos (\cite{fer12}), 
    radial profiles of well known radio halos have been modelled.
    
    In this work, we use three models of radio halo surface brightness taken from the literature. These models are described below.
    \begin{itemize}
        \item \textit{Gaussian:}\\
        The physical size of the radio halo is assumed to be the 
        FWHM (Full Width at Half Maximum) of a single Gaussian.
        The standard deviation of the model is then estimated from the FWHM.
        \item \textit{Polynomial:}\\
        A polynomial model of halo emission was first introduced by \cite{mjh17}
        where the authors fit the normalized brightness profile of 5 radio halos
        to a polynomial expression. They carried out a linear least square fit to the radial brightness distribution with  a quadratic function and obtained a best fit of the form: 
        \begin{equation}
            I(r) = -0.719r^2 + 1.867r - 0.141
        \end{equation}
        \item \textit{Exponential:}\\
        \cite{murgia09} found the brightness profile of radio halos to follow
        an exponential law in the form of:
        \begin{equation}
            I(r) = I_0e^{-r/r_e},
        \end{equation}
        where $r_e$ and $I_0$ are free parameters.
        \cite{bonafede17} fit the brightness profiles of 8 known halos and estimated
        $r_e = 2.6R_H$ where $R_H$ is the measured radius of the halo as estimated from images.
    \end{itemize}
    
    Figure \ref{fig:models} shows all three models for a fixed value of 
    $I_0$ (1 Jy), $R_H$ (1 Mpc) and $z$ (0.1).
    For our purposes, we have chosen the exponential model to be the default model
    used while creating model radio halos. We would like to  mention that while we have developed the tool with these three models, it is fairly straightforward to introduce a new model, should the need arise.
    
    In Figure~\ref{fig:model_profile} we also show the radial profiles of all three models discussed above. The profiles were calculated for the three images in Figure~\ref{fig:models}. Each pixel corresponds to $1.5"$ and the radius was calculated to be 139.33 pixels assuming $R_H$ = 1 Mpc and $z$ = 0.1.
    
\section{The Halo Upper Limit Calculator}\label{haloc}
    Radio halos being faint (surface brightnesses at 1.4 GHz typically $<1 \mu$Jy arcsec$^{-2}$)
    and extended (a few to several arcminutes depending on the redshift) sources, need high quality radio observations with radio interferometers
    to sample the angular scales and also to resolve the discrete sources such as point sources and radio galaxies. 
    The non-detection of such a source is also important to derive the implications to the physics of their origin
    (e.g. \cite{brunetti07}). The non-detection is a function of the sensitivity reached in the map and the uv-coverage on the angular scales where the emission is expected. Here, an empirical approach of injecting a model extended source in the visibilities and then checking for its detection is needed. 
    This process involves the steps of model preparation, injection of the model in the visibilities 
    and then making an image of the visibilities with the injected model. Once the image is made, the detection of the source needs to be checked for and this determines the direction of further model injection - whether a fainter or a brighter model is to be used. We describe the upper limit calculator that automates this process.
    
\begin{figure*}
        \centering
        \includegraphics[width=\textwidth]{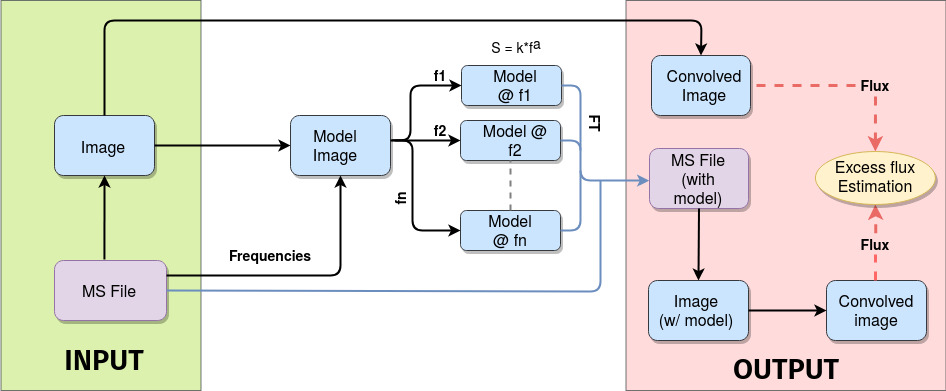}
        \caption{A schematic representation of the steps followed in the upper limit calculator.}
        \label{fig:ulc-flowchart}
    \end{figure*}

    The implementation is described using the case of determining upper limits for a radio halo in 
    a galaxy cluster. The steps are schematically shown in Fig.~\ref{fig:ulc-flowchart} and described 
    below.

    \begin{enumerate}
		\item The input files required are the final visibility file after self-calibration and the final image. The final image either already exists or is created as a first step.
		\item An image of the model radio halo of given flux, redshift, linear size and brightness profile is generated. The geometry of this image is identical to the image in the previous step. 
		\item The model image is Fourier transformed and added to the visibility file using standard CASA tasks.
		\item The new visibility file is imaged using the same parameters as used earlier to produce a new image.
		\item The difference in flux density around the central region between the image with and without the radio halo is estimated to calculate the ``recovered flux density''.
		\item The process is repeated for different values of halo flux densities keeping other parameters such as redshift, halo size fixed, and the new images are manually inspected to find the value at which a radio halo is visible. 
		\item This value can now be scaled to 1.4 GHz using an appropriate spectral index value to estimate the upper limit to the radio halo power. 
		This step is performed because most scaling relations for radio halos are determined using the radio power at 1.4 GHz (see for example \cite{wee19,fer12} for reviews).
	\end{enumerate}
	
	In the following subsections, the steps are described in further detail.
    All the tasks described below are performed using the Common Astronomy Software Applications
    (CASA) software (\cite{mcmullin07}) and makes use of the visibility
    and image architecture used by CASA.
        
    \subsection{Creating the model radio halo}
    The strength of the radio halo that needs to be injected in an image
    is entirely dependent on the quality of the image itself as well as the frequency
    of observation.
    The noise in the image can be used as a guide to decide on the flux densities of the halo
    that will be injected. 
    If the observation is also deep with good short spacing coverage, then the upper limit of halo emission will be correspondingly lower when compared to an image that is noisy.
    As a result, the first step of the process is to estimate the RMS in the image
    around the region where the source is expected to be.
    For our purposes, we are using the Background And Noise Estimation (BANE) program
    which comes packaged with the \href{https://github.com/PaulHancock/Aegean/wiki}{Aegean}\footnote{https://github.com/PaulHancock/Aegean/wiki} software. 
    
    The flux density of the injected radio halo is taken to be a factor times
    the calculated RMS in the image.
    Alternatively, a manual list of flux densities can also be provided to be injected.
    
    The only other physical property of the halo that is required is the angular size, $\theta$.
    This value depends on the redshift ($z$) of the target as well as the assumed physical size
    of the radio halo.
    Based on literature, the physical size of radio halos can be assumed to be $\sim 1$ Mpc and the angular size ($\theta$) subsequently estimated.
    This task uses the cosmology from \cite{planck15} with $H_0=67.7$, $\Omega_m=0.307$ and $\Omega_{total}=1$.
     
    The model of surface brightness distribution for the halo has been previously discussed  in Section~\ref{models}.
     
    \subsection{Injecting the radio halo}
    The radio halo created in the manner described above is only at a single frequency:
    the central bandwidth frequency of the input visibility file.
    Therefore we need to create additional images of the halo at all the other frequencies.
    
    This requires an assumption to be made regarding the frequency dependence of the radio halo
    flux density.
    The average spectral index ($\alpha$) of radio halos has been found to be $~-1.3$
    (\cite{fer12}). 
    
    We use this value to estimate the radio halo flux densities at other frequencies
    before taking Fourier transforms of the images and adding them to the input visibility file.
    
    \subsection{Estimating radio halo upper limit}
    Once all the halo visibilities have been added the next step is to image the new 
    visibility file, preferably to the same depth as the input image.
    Currently, we use the \texttt{tclean} task provided in CASA for imaging.
    
    The RMS estimated in the first step is used as a threshold during this
    deconvolution procedure. In order to better retrieve the extended emission in the image, we convolve the image with a wider beam to improve the signal to noise ratio.
    This can be done in multiple ways: a larger beam can be specified manually, or the beam can simply be a factor times the input beam, or the beam can be such that the halo surface area contains at least \texttt{nbeams} number of beams. 
    The tool uses the last option by default with \texttt{nbeams} as 100.

    \begin{figure*}
        \centering
        \includegraphics[width=\textwidth]{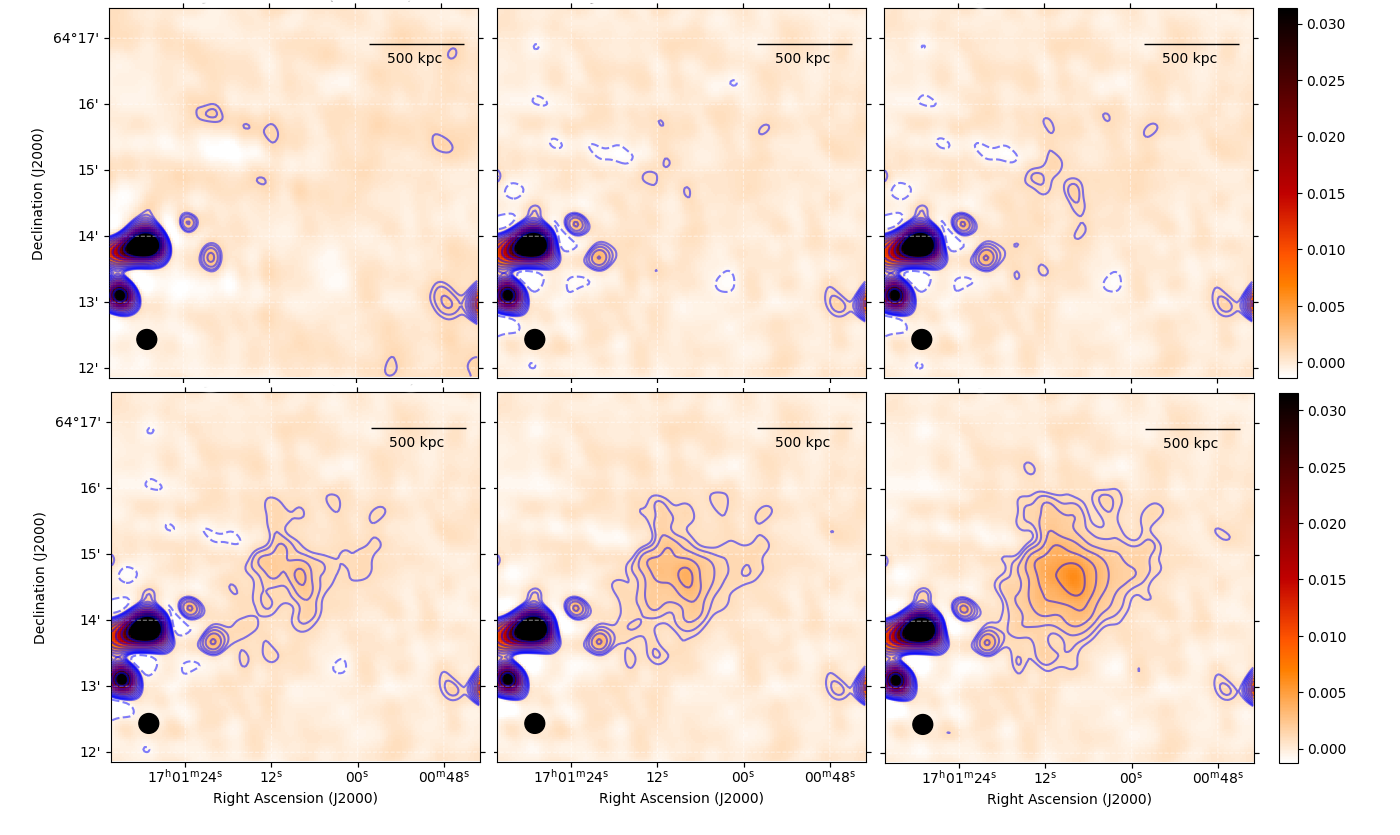}
        \caption{Progress of upper limit estimator for the cluster RXC J1701.3+6414. 
        Contours start at $3\sigma (\sigma=0.284$ mJy/beam) and increase by $\sqrt{2}$ afterwards.
        A single negative contour at $-3\sigma$ is also shown in dotted lines.
        The top left image is the original image created of the cluster.
        The next two images in the top row have halos with injected flux densities 14.21 and 28.41 mJy, respectively.
        The images in the bottom row (from left to right) have halos of injected flux densities 46.83, 85.24 and 142.07 mJy.
        Beam size in all images is the same and is shown in the bottom left corner
        ($\theta = 18''\times18''$).}
        \label{fig:combined}
    \end{figure*}
    
    Based on previous experience by the authors (\cite{kal13,kal15}) 
    most halo upper limits were calculated when the excess flux in the halo region was
    $\sim10\%$ higher than in the input image which has been our experience as well.
    \newline\newline
    The above steps are performed iteratively for different values of the input radio halo
    flux density until the excess flux exceeds $10\%$. 
    At this stage one still needs to manually inspect the images to deduce the flux density
    of the halo when it becomes ``visible''.
    
    The program allows the user to generate and view contours at each iteration of the process.
    By default contour levels start at $3\sigma$ (where $\sigma$ 
    is the RMS noise in the original image) 
    and increase by $\sqrt{2}$ afterwards. A single negative contour 
    at $-3\sigma$ is also plotted.
    
    \begin{table}
        \centering
        \scalebox{0.6}{
        \begin{tabular}{ll}
            Parameter name & Description \\
            \hline 
            \texttt{bane\_pth} & Path to BANE executable\\
            \texttt{srcdir}   &  Source Directory \\
            \texttt{visname}  & Reference visibility file \\
            \texttt{imgname}  & Reference image file made from 'visname'\\
            \texttt{cluster}  & Cluster name (optional)\\
            \texttt{z}        & Redshift of source $^{\dagger}$\\
            \texttt{l}        & Size of halo to be injected (kpc)\\
            \texttt{alpha}    & Spectral index for frequency scaling ($S \propto \nu^{-\alpha}$)\\
            \texttt{ftype}    & Radial profile of halo. Options: (G)aussian, (P)olynomial, (E)xponential\\
            \texttt{theta}    & Angular size (in arcsec) for halo (size=l) at redshift z $^{\dagger}$\\
            \texttt{x0, y0}   & Halo injection position\\
            \texttt{cell}     & Pixel separation (in arcsec)\\
            \texttt{hsize}    & Size of halo (in pixels) $^{\dagger}$\\
            \texttt{do\_fac}  & Whether calculate injection halo flux density or use manually (True or False)\\
            \texttt{flx\_fac}  & Flux level factors\\
            \texttt{flx\_lst} & Flux list (in Jy) \\
            \texttt{cln\_task} & Clean task to use ('tclean', 'wsclean') \\
            \texttt{N}        & No. of iterations\\      
            \texttt{isize}    & Image size $^{\dagger}$\\
            \texttt{csize}    & Cell size $^{\dagger}$\\   
            \texttt{weight}   & Weighting to be used\\  
            \texttt{dcv}      & Deconvolver to use\\     
            \texttt{scle}     & Multi-scale options\\    
            \texttt{thresh\_f} & Cleaning threshold factor\\
            \texttt{radius}   & Radius of halo (in arcsec) $^{\dagger}$\\  
            \texttt{rms\_reg}  & Region from which to estimate rms $^{\dagger}$\\ 
            \texttt{bopts}   & Method of image convolution ('num\_of\_beams', 'factor', 'beam')\\
            \texttt{nbeams}   & No. of synthesized beams in halo\\  
            \texttt{bparams}  & Beam size (bmaj("), bmin("), bpa(deg))\\ 
            \texttt{smooth\_f} & Factor to smooth input beam\\
            \texttt{recv\_th}  &  Threshold of Excess flux recovery at which to fine tune (in percent)\\
            \texttt{n\_split} & Number of levels to split during fine tuning \\
            \texttt{do\_contours} & Whether to make contours at the end of each iteration (True or False)\\
        \end{tabular}
        }
        \caption{Brief description of all options available in \texttt{params.py}. Parameters with the $\dagger$ superscript are automatically calculated based on other user provided paramters.}
        \label{tab:options}
    \end{table}{}
    
    \subsection{Options}
    Table \ref{tab:options} gives the list of various parameters that can be modified in
    \texttt{params.py}.
    Below we will briefly go through some of those options and their default values. 
    
    \begin{itemize}
        \item \texttt{l}: Giant radio halos are typically $\sim$1 Mpc in size.
        Default value: 1000 kpc
        \item \texttt{alpha}: Spectral index of the radio halo spectrum. Default value: $-1.3$
        \item \texttt{flx\_fac}: Factor times the RMS used to calculate injection halo flux density.
        Default values: [50, 100, 200, 300, 500, 1000, 2000]
        \item \texttt{cln\_task}: specifies which task should be used for deconvolution. Currently works best with CASA's
        \texttt{tclean} task.
        \item \texttt{thresh\_f}: Threshold to which to clean to. Default value: $3\sigma$
        \item \texttt{radius}: Calculated halo radius based on \texttt{l} and \texttt{z}
        \item \texttt{rms\_reg}: Radius of region in which to estimate RMS. 
        Default value: $3\times$\texttt{radius}
        \item \texttt{recv\_th}: Excess flux density recovery threshold. Default value: 10\%.
        \item \texttt{n\_split}: How many extra images to produce during fine-tuning (Default: 6)
    \end{itemize}
    
    \begin{figure*}
        \centering
        \includegraphics[width=\textwidth]{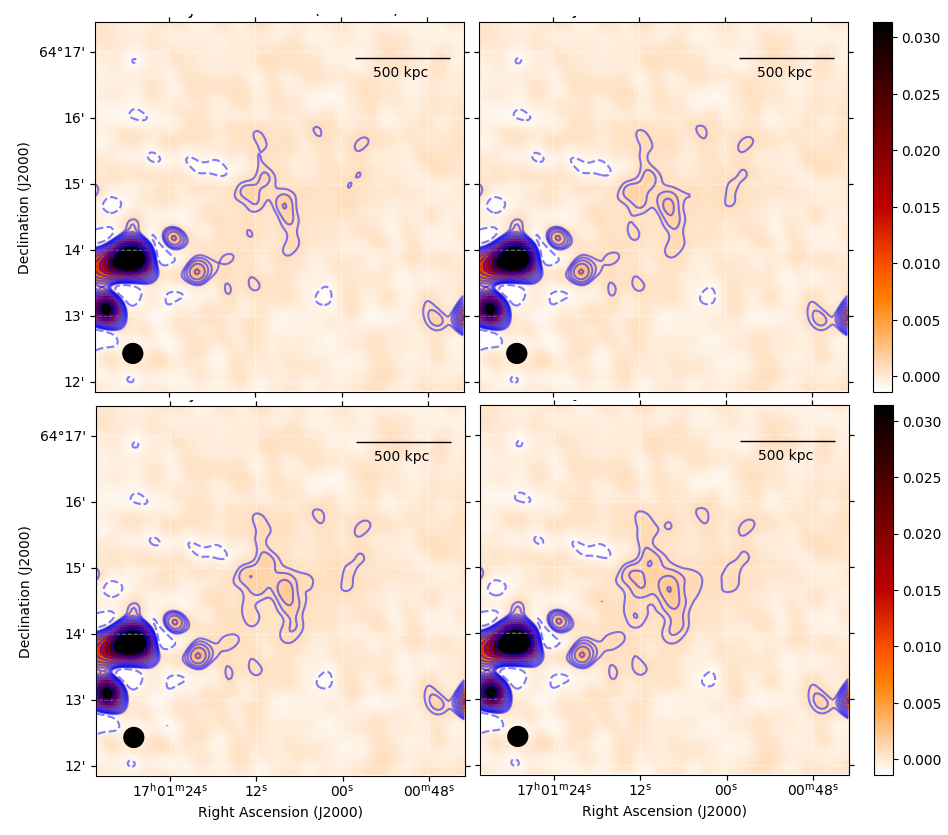}
        \caption{Sub images created between images 4 and 5 of Figure 3.
        Injected flux densities (left to right) are 34.1 and 39.78 mJy (top row) and 45.46 and 51.15 mJy (bottom row).}
        \label{fig:subiamges}
    \end{figure*}
    
\section{Illustration of using the UL calculator tool}\label{usage}
    Figure \ref{fig:combined} shows this process for the cluster RXC J1701.3+6414.
    The cluster was observed using the GMRT (Giant Metrewave Radio Telescope)
    on 21-08-2002 at 325 MHz for 189 minutes as part of the GMRT Key Project on galaxy clusters
    (George et al., in prep).
    The top left panel shows the original image (produced using the SPAM 
	(Source Peeling and Atmospheric Modeling) software (\cite{intema09,intema14})).
    The rest of the images have radio halos of various flux densities added to them 
    at the cluster centre.
    In the rest of the figures in the top panel the halo is still not visible.
    The figures in the bottom panel clearly detect the injected radio halo.
    The injected flux densities of the radio halos in the three images 
    (from left to right) are 56.83, 85.24 and 142.07 mJy.
    
	Quantitatively, this can also be seen through the contours seen in the figure.
	The RMS in the central region of the original image was estimated to be $0.284$ mJy/beam.
	We plot contours on the images starting at $3\sigma$ to trace the emergence of the radio halo in the subsequent images. 
	Plotting these contours provides another visual indication of the detection of a radio halo 
	of flux density 56.83 mJy.
	
	At this point fine-tuning can be performed between the flux range 28 and 56 mJy.
	Figure~\ref{fig:subiamges} shows the images of injected halos with flux densities
	between 28 and 56 mJy.
	Based on these images the injected halo with flux density 51.15 mJy 
	is the best image to claim detection of the radio halo. 
	 
	This can now be extrapolated to 1.4 GHz using 
	an average spectral index of $-1.3$ to get the radio halo upper limit at 1.4 GHz.
	Table~\ref{tab:recovery} also shows the injected flux densities and the recovered flux densities
	in the halo region for the cluster.
	
\section{Discussion}\label{disc}

    \begin{figure}
        \centering
        \includegraphics[width=\columnwidth]{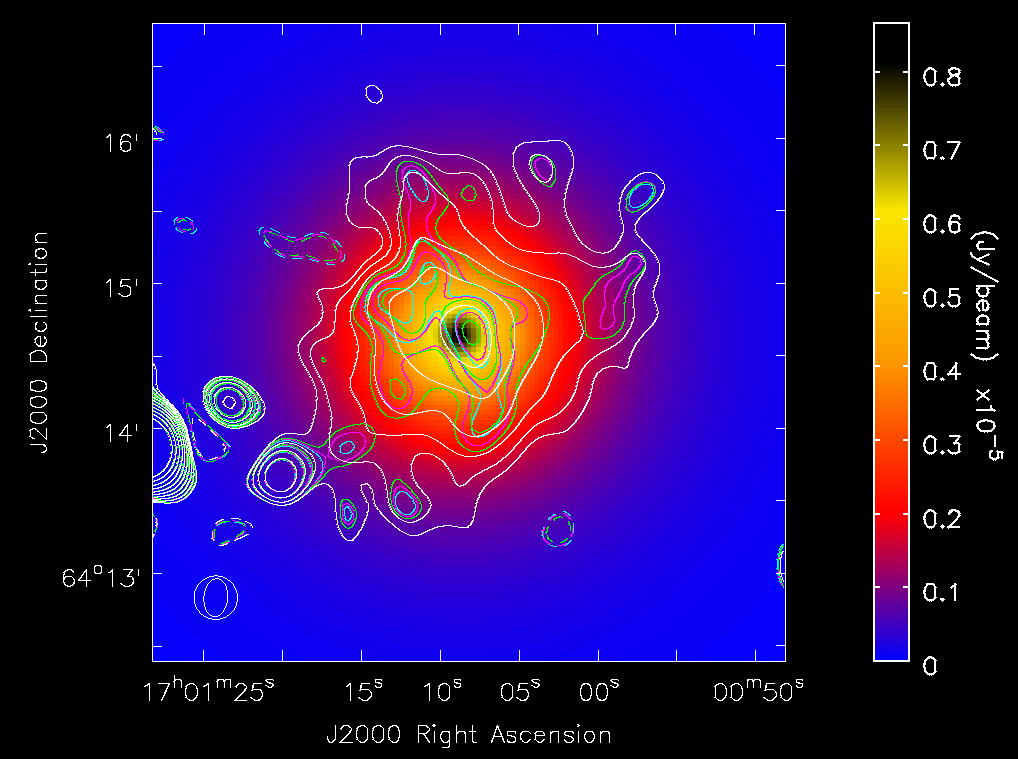}
        \caption{Image of the injected halo with flux density 14.21 mJy with contours of recovered halo of flux densities 34.1 mJy (cyan), 45.46 mJy (magenta), 56.83 mJy (green) and 142.07 mJy (white) plotted over.}
        \label{fig:overplot}
    \end{figure}
    
    \begin{figure}
        \centering
        \includegraphics[width=\columnwidth]{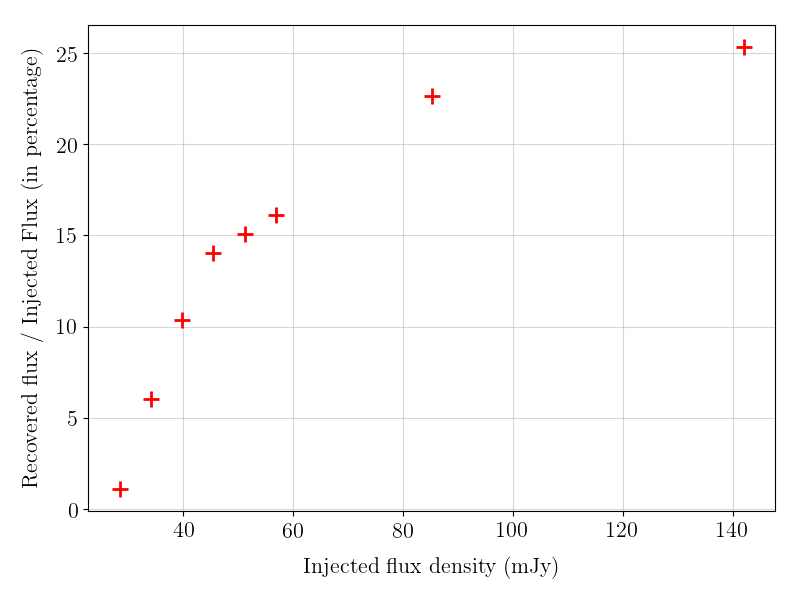}
        \caption{Plot of the fractional halo flux recovered (in percentage) vs injected flux for the cluster RXC J1701.3+6414.}
        \label{fig:recovered}
    \end{figure}
    
    \begin{table}
        \centering
        \begin{tabular}{ccc}
            Injected & Recovered & Percentage \\
            (mJy) & (mJy) & (\%)\\
            \hline
            14.21 & -4.68 & --    \\
            28.41 & 0.31  & 1.09  \\
            34.1  & 2.05  & 6.01  \\
            39.78 & 4.13  & 10.38 \\
            45.46 & 6.38  & 14.03 \\
            51.15 & 7.71  & 15.07 \\
            56.83 & 9.15  & 16.1  \\
            85.24 & 19.31 & 22.65 \\
            142.07& 36.02 & 25.35
        \end{tabular}
        \caption{Injected and recovered flux densities for the cluster RXC J1701.3+6414.
        All values are in mJy. Column 3 is the fractional flux density recovered.
        (Note that the negative flux recovered for the first injection is merely due to
        an improper deconvolution process and can be ignored.)}
        \label{tab:recovery}
    \end{table}
    
    The main advantage of this tool is the control provided to the user to inject
    halos of various configurations and subsequently estimate upper limits in a
    quick and automated manner.
    The software can also be used to estimate upper limits to emission of
    other extended radio sources such as mini-halos.
    This is because the models of their emission can also be approximated to
    one of the three models we have used to model radio halos (\cite{murgia09,basu16}).
    In practice, the tool has been used to estimate upper limits for 20 clusters 
    which were part of the sample in the GMRT Key Project on Galaxy clusters
    (George et al. in preparation). 
    
    \subsection{Uncertainty on the obtained upper limit}
    It should be noted that at the current stage the process for upper limit estimation
    is still dependent on human intervention at the final step.
    While most of the steps of this analysis are automated, confirmation of detection
    still requires manual input.
    This is because a robust detection of a radio halo in a cluster is dependent 
    on a lot of factors.
    These include the depth of the observations, the quality of the image
    and even the local environment around the central regions of the cluster.
    As a result manual inspection on this final step is still recommended
    to confirm the ``appearance'' of an injected radio halo.
    The recovery threshold value of 10\% is still accurate in terms of estimating
    the approximate range of flux density values at which the radio halo appears.
    
    Figure~\ref{fig:overplot} shows an image of a model halo with exponential profile and flux density 14.21 mJy. 
    Plotted on top of the image are contours of recovered halos with flux densities 34.1 mJy (cyan), 45.46 mJy (magenta), 56.83 mJy (green) and 142.07 mJy (white). 
    These are the same contours as shown in Figures~\ref{fig:combined} and \ref{fig:subiamges}. 
    This shows that the recovered halo image is smaller in size compared to the injected halo until it crosses a certain flux density threshold.
    
    In Figure \ref{fig:recovered} we have plotted the fractional flux ``recovered'' vs the
    injected flux density of the halo for the cluster RXC J1701.3+6414.
    The knee of the plot occurs when the injected flux density is $40-50$ mJy.
    This is in agreement with what we have seen from Figures~\ref{fig:combined} and \ref{fig:subiamges}.
    We believe that the 10\% recovery threshold chosen is a reasonable
    estimator to confirm the detection limit of the halo and fine-tuning around that range
    can provide a more accurate estimate of the upper limit.
    Please note that first data point from Table~\ref{tab:recovery} is not plotted in 
    the figure for the sake of clarity.
    
    \cite{bonafede17} also performed a similar analysis on a sample of clusters but
    used a different measure to confirm the detection of an injected halo.
    A halo was considered to be detected if the $2\sigma$ size of the source ($D_{2\sigma}$)
    was greater than half the size of the injected halo, or ($R_H$).
    However, they did not inject their halos in the central regions of the cluster,
    instead choosing a location near the cluster, free from any confusion due to 
    unresolved sources. 
    We do not believe this is an accurate estimation for a cluster's upper limit.
    Cluster radio halos are almost always found in the central regions of the cluster
    and have to contend with the presence of existing sources.
    This could be unresolved radio sources such as the BCGs (Brightest Cluster Galaxies) 
    of the cluster or even tailed radio galaxies (e.g. A2163, \cite{feretti01}).
    In order to get the best estimate of the halo upper limit, we believe that the artificial halo
    should be injected at the cluster centre and the total flux density in the new image
    compared with the original image.
    
\subsection{Application to all sky surveys}
    Large area sky surveys such as the NVSS (NRAO Very large Sky Survey, \cite{con98})
    were crucial in the discovery of the earliest samples of cluster radio halos and
    relics (\cite{giovannini99}).
    This software can also be used to estimate average detection limits for the entire survey
    by selecting representative fields in the survey and estimating the upper limits
    in those surveys.
    Upper limits calculated in such a manner can represent the detection limit for extended sources in those surveys.
    
    This will have implications to next generation radio surveys like the LOFAR
    Two-metre Sky Survey (LoTSS, \cite{shimwell17,shimwell19}) 
    and surveys in the the future using the Square Kilometre Array (SKA).
    The sensitivity that these surveys are expected to reach is below those of current generation radio telescopes (e.g. $<100 \mu$Jy at 150 MHz for LoTSS).
    These surveys are expected to result in the detection of many more new sources of
    diffuse extended emission (\cite{bonafede15,mjh15}) including previously undetected
    radio halos.
    Radio halo detections as well as non-detections from these next generation of telescopes will have implications on the physics of their origin. (\cite{brunetti07,brulaz16}).

\section{Conclusions}\label{conc}
    The vast majority of galaxy clusters do not contain detectable radio halos; diffuse and extended
    sources of synchrotron radio emission produced in the intra-cluster gas.
    It is unclear if the lack of detection is due to a genuine absence of emission
    or if the emission is too weak to be detected by the current generation of telescopes.
    However, it is still possible to estimate upper limits to a possible radio emission.
    These upper limits can help put constraints on the various competing models of halo emission.
    
    In this paper, we have presented a Python based tool that can be used to 
    automate the process of estimating upper limits to radio halo emission.
    A user defined model of a radio halo is generated, Fourier transformed and then injected 
    into the input visibility file and imaged.
    The halo injected image is then compared to the original image to detect the presence of a radio halo.
    On the basis of this comparison, the software decides either to increase or decrease the 
    injection flux density so as to best reach the true ``upper limit'' of detection.
    At every step, an image can also be produced for the user to check the progress of the analysis.
    
    This tool has been successfully tested on a large number of galaxy cluster observations
    with reasonable results (George et al. in preparation).
    Currently, the tool is optimised for the injection and detection of radio halos
    although other extended sources such as mini-halos, radio relics and supernovae remnants
    could also be modelled and their upper limits estimated.
    The python code for estimating the upper limits can be found
    \href{https://github.com/lijotgeorge/UL-CALC}{here}\footnote{https://github.com/lijotgeorge/UL-CALC}. 

\begin{acknowledgements}
We acknowledge the support of the Department of Atomic Energy, Government of
India, under project no. 12-R\&D-TFR-5.02-0700. We thank the staff of the GMRT that made these observations possible. GMRT is run by the National Centre for Radio Astrophysics of the Tata Institute of Fundamental Research. RK acknowledges support from the DST-INSPIRE Faculty Award of the Government of India.
This research made use of Astropy,\footnote{http://www.astropy.org} a community-developed core Python package for Astronomy (\cite{astropy:2013,astropy:2018}).
This research made use of Astroquery, an astropy affiliated package that contains a collection of tools to access online Astronomical data (\cite{astroquery}).
We would also like to thank the anonymous referee for their comments and suggestions that have helped improve this paper.
\end{acknowledgements}

%%%%%%%%%%%%%%%%%%%%%%%%%%%%%%%%%%%%%%%%%%%%%%%%%%



% \bibliographystyle{mnras}
\bibliographystyle{apalike}
\bibliography{all_ref} % if your bibtex file is called example.bib

\begin{thebibliography}{}

\bibitem[{Astropy Collaboration} et~al., 2013]{astropy:2013}
{Astropy Collaboration}, {Robitaille}, T.~P., {Tollerud}, E.~J., {Greenfield},
  P., {Droettboom}, M., {Bray}, E., {Aldcroft}, T., {Davis}, M., {Ginsburg},
  A., {Price-Whelan}, A.~M., {Kerzendorf}, W.~E., {Conley}, A., {Crighton}, N.,
  {Barbary}, K., {Muna}, D., {Ferguson}, H., {Grollier}, F., {Parikh}, M.~M.,
  {Nair}, P.~H., {Unther}, H.~M., {Deil}, C., {Woillez}, J., {Conseil}, S.,
  {Kramer}, R., {Turner}, J.~E.~H., {Singer}, L., {Fox}, R., {Weaver}, B.~A.,
  {Zabalza}, V., {Edwards}, Z.~I., {Azalee Bostroem}, K., {Burke}, D.~J.,
  {Casey}, A.~R., {Crawford}, S.~M., {Dencheva}, N., {Ely}, J., {Jenness}, T.,
  {Labrie}, K., {Lim}, P.~L., {Pierfederici}, F., {Pontzen}, A., {Ptak}, A.,
  {Refsdal}, B., {Servillat}, M., and {Streicher}, O. (2013).
\newblock {Astropy: A community Python package for astronomy}.
\newblock {\em \aap}, 558:A33.

\bibitem[{Basu} et~al., 2016]{basu16}
{Basu}, K., {Vazza}, F., {Erler}, J., and {Sommer}, M. (2016).
\newblock {The impact of the SZ effect on cm-wavelength (1-30 GHz) observations
  of galaxy cluster radio relics}.
\newblock {\em \aap}, 591:A142.

\bibitem[{Bonafede} et~al., 2017]{bonafede17}
{Bonafede}, A., {Cassano}, R., {Br{\"u}ggen}, M., {Ogrean}, G.~A., {Riseley},
  C.~J., {Cuciti}, V., {de Gasperin}, F., {Golovich}, N., {Kale}, R.,
  {Venturi}, T., {van Weeren}, R.~J., {Wik}, D.~R., and {Wittman}, D. (2017).
\newblock {On the absence of radio haloes in clusters with double relics}.
\newblock {\em Monthly Notices of the Royal Astronomical Society},
  470(3):3465--3475.

\bibitem[{Bonafede} et~al., 2015]{bonafede15}
{Bonafede}, A., {Vazza}, F., {Br{\"u}ggen}, M., {Akahori}, T., {Carretti}, E.,
  {Colafrancesco}, S., {Feretti}, L., {Ferrari}, C., {Giovannini}, G.,
  {Govoni}, F., {Johnston-Hollitt}, M., {Murgia}, M., {Scaife}, A., {Vacca},
  V., {Govoni}, F., {Rudnick}, L., and {Scaife}, A. (2015).
\newblock {Unravelling the origin of large-scale magnetic fields in galaxy
  clusters and beyond through Faraday Rotation Measures with the SKA}.
\newblock In {\em Advancing Astrophysics with the Square Kilometre Array
  (AASKA14)}, page~95.

\bibitem[{Brunetti} and {Lazarian}, 2016]{brulaz16}
{Brunetti}, G. and {Lazarian}, A. (2016).
\newblock {Stochastic reacceleration of relativistic electrons by turbulent
  reconnection: a mechanism for cluster-scale radio emission?}
\newblock {\em \mnras}, 458:2584--2595.

\bibitem[{Brunetti} et~al., 2007]{brunetti07}
{Brunetti}, G., {Venturi}, T., {Dallacasa}, D., {Cassano}, R., {Dolag}, K.,
  {Giacintucci}, S., and {Setti}, G. (2007).
\newblock {Cosmic Rays and Radio Halos in Galaxy Clusters: New Constraints from
  Radio Observations}.
\newblock {\em \apjl}, 670:L5--L8.

\bibitem[{Condon} et~al., 1998]{con98}
{Condon}, J.~J., {Cotton}, W.~D., {Greisen}, E.~W., {Yin}, Q.~F., {Perley},
  R.~A., {Taylor}, G.~B., and {Broderick}, J.~J. (1998).
\newblock {The NRAO VLA Sky Survey}.
\newblock {\em AJ}, 115:1693--1716.

\bibitem[{Deo} and {Kale}, 2017]{deo17}
{Deo}, D.~K. and {Kale}, R. (2017).
\newblock {Simulations of imaging extended sources using the GMRT and the
  U-GMRT. Implications to observing strategies}.
\newblock {\em Experimental Astronomy}, 44:165--180.

\bibitem[{Feretti} et~al., 2001]{feretti01}
{Feretti}, L., {Fusco-Femiano}, R., {Giovannini}, G., and {Govoni}, F. (2001).
\newblock {The giant radio halo in Abell 2163}.
\newblock {\em \aap}, 373:106--112.

\bibitem[{Feretti} et~al., 2012]{fer12}
{Feretti}, L., {Giovannini}, G., {Govoni}, F., and {Murgia}, M. (2012).
\newblock {Clusters of galaxies: observational properties of the diffuse radio
  emission}.
\newblock {\em AApR}, 20:54.

\bibitem[{Ginsburg} et~al., 2019]{astroquery}
{Ginsburg}, A., {Sip{\\H o}cz}, B.~M., {Brasseur}, C.~E., {Cowperthwaite},
  P.~S., {Craig}, M.~W., {Deil}, C., {Guillochon}, J., {Guzman}, G., {Liedtke},
  S., {Lian Lim}, P., {Lockhart}, K.~E., {Mommert}, M., {Morris}, B.~M.,
  {Norman}, H., {Parikh}, M., {Persson}, M.~V., {Robitaille}, T.~P., {Segovia},
  J.-C., {Singer}, L.~P., {Tollerud}, E.~J., {de Val-Borro}, M., {Valtchanov},
  I., {Woillez}, J., {The Astroquery collaboration}, and {a subset of the
  astropy collaboration} (2019).
\newblock {astroquery: An Astronomical Web-querying Package in Python}.
\newblock {\em \\aj}, 157:98.

\bibitem[{Giovannini} et~al., 1999]{giovannini99}
{Giovannini}, G., {Tordi}, M., and {Feretti}, L. (1999).
\newblock {Radio halo and relic candidates from the NRAO VLA Sky Survey}.
\newblock {\em \na}, 4:141--155.

\bibitem[{Intema}, 2014]{intema14}
{Intema}, H.~T. (2014).
\newblock {SPAM: A data reduction recipe for high-resolution,low-frequency
  radio-interferometric observations}.
\newblock In {\em Astronomical Society of India Conference Series}, volume~13
  of {\em Astronomical Society of India Conference Series}.

\bibitem[{Intema} et~al., 2009]{intema09}
{Intema}, H.~T., {van der Tol}, S., {Cotton}, W.~D., {Cohen}, A.~S., {van
  Bemmel}, I.~M., and {R{\"o}ttgering}, H.~J.~A. (2009).
\newblock {Ionospheric calibration of low frequency radio interferometric
  observations using the peeling scheme. I. Method description and first
  results}.
\newblock {\em \aap}, 501:1185--1205.

\bibitem[{Johnston-Hollitt} et~al., 2015]{mjh15}
{Johnston-Hollitt}, M., {Govoni}, F., {Beck}, R., {Dehghan}, S., {Pratley}, L.,
  {Akahori}, T., {Heald}, G., {Agudo}, I., {Bonafede}, A., {Carretti}, E.,
  {Clarke}, T., {Colafrancesco}, S., {Ensslin}, T.~A., {Feretti}, L.,
  {Gaensler}, B., {Haverkorn}, M., {Mao}, S.~A., {Oppermann}, N., {Rudnick},
  L., {Scaife}, A., {Schnitzeler}, D., {Stil}, J., {Taylor}, A.~R., and
  {Vacca}, V. (2015).
\newblock {Using SKA Rotation Measures to Reveal the Mysteries of the
  Magnetised Universe}.
\newblock In {\em Advancing Astrophysics with the Square Kilometre Array
  (AASKA14)}, page~92.

\bibitem[{Johnston-Hollitt} and {Pratley}, 2017]{mjh17}
{Johnston-Hollitt}, M. and {Pratley}, L. (2017).
\newblock {Upper limits on a radio halo in Abell 3667 at 1.4 GHz}.
\newblock {\em arXiv e-prints}, page arXiv:1706.04930.

\bibitem[{Kale} et~al., 2015]{kal15}
{Kale}, R., {Venturi}, T., {Giacintucci}, S., {Dallacasa}, D., {Cassano}, R.,
  {Brunetti}, G., {Cuciti}, V., {Macario}, G., and {Athreya}, R. (2015).
\newblock {The Extended GMRT Radio Halo Survey. II. Further results and
  analysis of the full sample}.
\newblock {\em A\&A}, 579:A92.

\bibitem[{Kale} et~al., 2013]{kal13}
{Kale}, R., {Venturi}, T., {Giacintucci}, S., {Dallacasa}, D., {Cassano}, R.,
  {Brunetti}, G., {Macario}, G., and {Athreya}, R. (2013).
\newblock {The Extended GMRT Radio Halo Survey. I. New upper limits on radio
  halos and mini-halos}.
\newblock {\em A\&A}, 557:A99.

\bibitem[{McMullin} et~al., 2007]{mcmullin07}
{McMullin}, J.~P., {Waters}, B., {Schiebel}, D., {Young}, W., and {Golap}, K.
  (2007).
\newblock {\em {CASA Architecture and Applications}}, volume 376 of {\em
  Astronomical Society of the Pacific Conference Series}, page 127.

\bibitem[{Murgia} et~al., 2009]{murgia09}
{Murgia}, M., {Govoni}, F., {Markevitch}, M., {Feretti}, L., {Giovannini}, G.,
  {Taylor}, G.~B., and {Carretti}, E. (2009).
\newblock {Comparative analysis of the diffuse radio emission in the galaxy
  clusters A1835, A2029, and Ophiuchus}.
\newblock {\em Astronomy and Astrophysics}, 499(3):679--695.

\bibitem[{Nayana} et~al., 2017]{nayana17}
{Nayana}, A.~J., {Chandra}, P., {Roy}, S., {Green}, D.~A., {Acero}, F.,
  {Lemoine-Goumard}, M., {Marcowith}, A., {Ray}, A.~K., and {Renaud}, M.
  (2017).
\newblock {325 and 610 MHz radio counterparts of SNR G353.6-0.7 also known as
  HESS J1731-347}.
\newblock {\em \mnras}, 467(1):155--163.

\bibitem[{Planck Collaboration} et~al., 2016]{planck15}
{Planck Collaboration}, {Ade}, P.~A.~R., {Aghanim}, N., {Arnaud}, M.,
  {Ashdown}, M., {Aumont}, J., {Baccigalupi}, C., {Banday}, A.~J., {Barreiro},
  R.~B., {Bartlett}, J.~G., {Bartolo}, N., {Battaner}, E., {Battye}, R.,
  {Benabed}, K., {Beno{\^\i}t}, A., {Benoit-L{\'e}vy}, A., {Bernard}, J.~P.,
  {Bersanelli}, M., {Bielewicz}, P., {Bock}, J.~J., {Bonaldi}, A., {Bonavera},
  L., {Bond}, J.~R., {Borrill}, J., {Bouchet}, F.~R., {Boulanger}, F.,
  {Bucher}, M., {Burigana}, C., {Butler}, R.~C., {Calabrese}, E., {Cardoso},
  J.~F., {Catalano}, A., {Challinor}, A., {Chamballu}, A., {Chary}, R.~R.,
  {Chiang}, H.~C., {Chluba}, J., {Christensen}, P.~R., {Church}, S.,
  {Clements}, D.~L., {Colombi}, S., {Colombo}, L.~P.~L., {Combet}, C.,
  {Coulais}, A., {Crill}, B.~P., {Curto}, A., {Cuttaia}, F., {Danese}, L.,
  {Davies}, R.~D., {Davis}, R.~J., {de Bernardis}, P., {de Rosa}, A., {de
  Zotti}, G., {Delabrouille}, J., {D{\'e}sert}, F.~X., {Di Valentino}, E.,
  {Dickinson}, C., {Diego}, J.~M., {Dolag}, K., {Dole}, H., {Donzelli}, S.,
  {Dor{\'e}}, O., {Douspis}, M., {Ducout}, A., {Dunkley}, J., {Dupac}, X.,
  {Efstathiou}, G., {Elsner}, F., {En{\ss}lin}, T.~A., {Eriksen}, H.~K.,
  {Farhang}, M., {Fergusson}, J., {Finelli}, F., {Forni}, O., {Frailis}, M.,
  {Fraisse}, A.~A., {Franceschi}, E., {Frejsel}, A., {Galeotta}, S., {Galli},
  S., {Ganga}, K., {Gauthier}, C., {Gerbino}, M., {Ghosh}, T., {Giard}, M.,
  {Giraud-H{\'e}raud}, Y., {Giusarma}, E., {Gjerl{\o}w}, E.,
  {Gonz{\'a}lez-Nuevo}, J., {G{\'o}rski}, K.~M., {Gratton}, S., {Gregorio}, A.,
  {Gruppuso}, A., {Gudmundsson}, J.~E., {Hamann}, J., {Hansen}, F.~K.,
  {Hanson}, D., {Harrison}, D.~L., {Helou}, G., {Henrot-Versill{\'e}}, S.,
  {Hern{\'a}ndez-Monteagudo}, C., {Herranz}, D., {Hildebrand t}, S.~R.,
  {Hivon}, E., {Hobson}, M., {Holmes}, W.~A., {Hornstrup}, A., {Hovest}, W.,
  {Huang}, Z., {Huffenberger}, K.~M., {Hurier}, G., {Jaffe}, A.~H., {Jaffe},
  T.~R., {Jones}, W.~C., {Juvela}, M., {Keih{\"a}nen}, E., {Keskitalo}, R.,
  {Kisner}, T.~S., {Kneissl}, R., {Knoche}, J., {Knox}, L., {Kunz}, M.,
  {Kurki-Suonio}, H., {Lagache}, G., {L{\"a}hteenm{\"a}ki}, A., {Lamarre},
  J.~M., {Lasenby}, A., {Lattanzi}, M., {Lawrence}, C.~R., {Leahy}, J.~P.,
  {Leonardi}, R., {Lesgourgues}, J., {Levrier}, F., {Lewis}, A., {Liguori}, M.,
  {Lilje}, P.~B., {Linden-V{\o}rnle}, M., {L{\'o}pez-Caniego}, M., {Lubin},
  P.~M., {Mac{\'\i}as-P{\'e}rez}, J.~F., {Maggio}, G., {Maino}, D.,
  {Mandolesi}, N., {Mangilli}, A., {Marchini}, A., {Maris}, M., {Martin},
  P.~G., {Martinelli}, M., {Mart{\'\i}nez-Gonz{\'a}lez}, E., {Masi}, S.,
  {Matarrese}, S., {McGehee}, P., {Meinhold}, P.~R., {Melchiorri}, A., {Melin},
  J.~B., {Mendes}, L., {Mennella}, A., {Migliaccio}, M., {Millea}, M., {Mitra},
  S., {Miville-Desch{\^e}nes}, M.~A., {Moneti}, A., {Montier}, L., {Morgante},
  G., {Mortlock}, D., {Moss}, A., {Munshi}, D., {Murphy}, J.~A., {Naselsky},
  P., {Nati}, F., {Natoli}, P., {Netterfield}, C.~B., {N{\o}rgaard-Nielsen},
  H.~U., {Noviello}, F., {Novikov}, D., {Novikov}, I., {Oxborrow}, C.~A.,
  {Paci}, F., {Pagano}, L., {Pajot}, F., {Paladini}, R., {Paoletti}, D.,
  {Partridge}, B., {Pasian}, F., {Patanchon}, G., {Pearson}, T.~J.,
  {Perdereau}, O., {Perotto}, L., {Perrotta}, F., {Pettorino}, V.,
  {Piacentini}, F., {Piat}, M., {Pierpaoli}, E., {Pietrobon}, D.,
  {Plaszczynski}, S., {Pointecouteau}, E., {Polenta}, G., {Popa}, L., {Pratt},
  G.~W., {Pr{\'e}zeau}, G., {Prunet}, S., {Puget}, J.~L., {Rachen}, J.~P.,
  {Reach}, W.~T., {Rebolo}, R., {Reinecke}, M., {Remazeilles}, M., {Renault},
  C., {Renzi}, A., {Ristorcelli}, I., {Rocha}, G., {Rosset}, C., {Rossetti},
  M., {Roudier}, G., {Rouill{\'e} d'Orfeuil}, B., {Rowan-Robinson}, M.,
  {Rubi{\~n}o-Mart{\'\i}n}, J.~A., {Rusholme}, B., {Said}, N., {Salvatelli},
  V., {Salvati}, L., {Sandri}, M., {Santos}, D., {Savelainen}, M., {Savini},
  G., {Scott}, D., {Seiffert}, M.~D., {Serra}, P., {Shellard}, E.~P.~S.,
  {Spencer}, L.~D., {Spinelli}, M., {Stolyarov}, V., {Stompor}, R., {Sudiwala},
  R., {Sunyaev}, R., {Sutton}, D., {Suur-Uski}, A.~S., {Sygnet}, J.~F.,
  {Tauber}, J.~A., {Terenzi}, L., {Toffolatti}, L., {Tomasi}, M., {Tristram},
  M., {Trombetti}, T., {Tucci}, M., {Tuovinen}, J., {T{\"u}rler}, M., {Umana},
  G., {Valenziano}, L., {Valiviita}, J., {Van Tent}, F., {Vielva}, P., {Villa},
  F., {Wade}, L.~A., {Wandelt}, B.~D., {Wehus}, I.~K., {White}, M., {White},
  S.~D.~M., {Wilkinson}, A., {Yvon}, D., {Zacchei}, A., and {Zonca}, A. (2016).
\newblock {Planck 2015 results. XIII. Cosmological parameters}.
\newblock {\em \aap}, 594:A13.

\bibitem[{Price-Whelan} et~al., 2018]{astropy:2018}
{Price-Whelan}, A.~M., {Sip{\H{o}}cz}, B.~M., {G{\"u}nther}, H.~M., {Lim},
  P.~L., {Crawford}, S.~M., {Conseil}, S., {Shupe}, D.~L., {Craig}, M.~W.,
  {Dencheva}, N., {Ginsburg}, A., {VanderPlas}, J.~T., {Bradley}, L.~D.,
  {P{\'e}rez-Su{\'a}rez}, D., {de Val-Borro}, M., {Paper Contributors}, P.,
  {Aldcroft}, T.~L., {Cruz}, K.~L., {Robitaille}, T.~P., {Tollerud}, E.~J.,
  {Coordination Committee}, A., {Ardelean}, C., {Babej}, T., {Bach}, Y.~P.,
  {Bachetti}, M., {Bakanov}, A.~V., {Bamford}, S.~P., {Barentsen}, G.,
  {Barmby}, P., {Baumbach}, A., {Berry}, K.~L., {Biscani}, F., {Boquien}, M.,
  {Bostroem}, K.~A., {Bouma}, L.~G., {Brammer}, G.~B., {Bray}, E.~M.,
  {Breytenbach}, H., {Buddelmeijer}, H., {Burke}, D.~J., {Calderone}, G., {Cano
  Rodr{\'\i}guez}, J.~L., {Cara}, M., {Cardoso}, J.~V.~M., {Cheedella}, S.,
  {Copin}, Y., {Corrales}, L., {Crichton}, D., {D{\textquoteright}Avella}, D.,
  {Deil}, C., {Depagne}, {\'E}., {Dietrich}, J.~P., {Donath}, A., {Droettboom},
  M., {Earl}, N., {Erben}, T., {Fabbro}, S., {Ferreira}, L.~A., {Finethy}, T.,
  {Fox}, R.~T., {Garrison}, L.~H., {Gibbons}, S.~L.~J., {Goldstein}, D.~A.,
  {Gommers}, R., {Greco}, J.~P., {Greenfield}, P., {Groener}, A.~M.,
  {Grollier}, F., {Hagen}, A., {Hirst}, P., {Homeier}, D., {Horton}, A.~J.,
  {Hosseinzadeh}, G., {Hu}, L., {Hunkeler}, J.~S., {Ivezi{\'c}}, {\v{Z}}.,
  {Jain}, A., {Jenness}, T., {Kanarek}, G., {Kendrew}, S., {Kern}, N.~S.,
  {Kerzendorf}, W.~E., {Khvalko}, A., {King}, J., {Kirkby}, D., {Kulkarni},
  A.~M., {Kumar}, A., {Lee}, A., {Lenz}, D., {Littlefair}, S.~P., {Ma}, Z.,
  {Macleod}, D.~M., {Mastropietro}, M., {McCully}, C., {Montagnac}, S.,
  {Morris}, B.~M., {Mueller}, M., {Mumford}, S.~J., {Muna}, D., {Murphy},
  N.~A., {Nelson}, S., {Nguyen}, G.~H., {Ninan}, J.~P., {N{\"o}the}, M.,
  {Ogaz}, S., {Oh}, S., {Parejko}, J.~K., {Parley}, N., {Pascual}, S., {Patil},
  R., {Patil}, A.~A., {Plunkett}, A.~L., {Prochaska}, J.~X., {Rastogi}, T.,
  {Reddy Janga}, V., {Sabater}, J., {Sakurikar}, P., {Seifert}, M., {Sherbert},
  L.~E., {Sherwood-Taylor}, H., {Shih}, A.~Y., {Sick}, J., {Silbiger}, M.~T.,
  {Singanamalla}, S., {Singer}, L.~P., {Sladen}, P.~H., {Sooley}, K.~A.,
  {Sornarajah}, S., {Streicher}, O., {Teuben}, P., {Thomas}, S.~W., {Tremblay},
  G.~R., {Turner}, J.~E.~H., {Terr{\'o}n}, V., {van Kerkwijk}, M.~H., {de la
  Vega}, A., {Watkins}, L.~L., {Weaver}, B.~A., {Whitmore}, J.~B., {Woillez},
  J., {Zabalza}, V., and {Contributors}, A. (2018).
\newblock {The Astropy Project: Building an Open-science Project and Status of
  the v2.0 Core Package}.
\newblock {\em \aj}, 156:123.

\bibitem[{Shimwell} et~al., 2017]{shimwell17}
{Shimwell}, T.~W., {R{\"o}ttgering}, H.~J.~A., {Best}, P.~N., {Williams},
  W.~L., {Dijkema}, T.~J., {de Gasperin}, F., {Hardcastle}, M.~J., {Heald},
  G.~H., {Hoang}, D.~N., {Horneffer}, A., {Intema}, H., {Mahony}, E.~K.,
  {Mandal}, S., {Mechev}, A.~P., {Morabito}, L., {Oonk}, J.~B.~R., {Rafferty},
  D., {Retana-Montenegro}, E., {Sabater}, J., {Tasse}, C., {van Weeren}, R.~J.,
  {Br{\"u}ggen}, M., {Brunetti}, G., {Chy{\.z}y}, K.~T., {Conway}, J.~E.,
  {Haverkorn}, M., {Jackson}, N., {Jarvis}, M.~J., {McKean}, J.~P., {Miley},
  G.~K., {Morganti}, R., {White}, G.~J., {Wise}, M.~W., {van Bemmel}, I.~M.,
  {Beck}, R., {Brienza}, M., {Bonafede}, A., {Calistro Rivera}, G., {Cassano},
  R., {Clarke}, A.~O., {Cseh}, D., {Deller}, A., {Drabent}, A., {van Driel},
  W., {Engels}, D., {Falcke}, H., {Ferrari}, C., {Fr{\"o}hlich}, S., {Garrett},
  M.~A., {Harwood}, J.~J., {Heesen}, V., {Hoeft}, M., {Horellou}, C., {Israel},
  F.~P., {Kapi{\'n}ska}, A.~D., {Kunert-Bajraszewska}, M., {McKay}, D.~J.,
  {Mohan}, N.~R., {Orr{\'u}}, E., {Pizzo}, R.~F., {Prandoni}, I., {Schwarz},
  D.~J., {Shulevski}, A., {Sipior}, M., {Smith}, D.~J.~B., {Sridhar}, S.~S.,
  {Steinmetz}, M., {Stroe}, A., {Varenius}, E., {van der Werf}, P.~P.,
  {Zensus}, J.~A., and {Zwart}, J.~T.~L. (2017).
\newblock {The LOFAR Two-metre Sky Survey. I. Survey description and
  preliminary data release}.
\newblock {\em \aap}, 598:A104.

\bibitem[{Shimwell} et~al., 2019]{shimwell19}
{Shimwell}, T.~W., {Tasse}, C., {Hardcastle}, M.~J., {Mechev}, A.~P.,
  {Williams}, W.~L., {Best}, P.~N., {R{\"o}ttgering}, H.~J.~A., {Callingham},
  J.~R., {Dijkema}, T.~J., {de Gasperin}, F., {Hoang}, D.~N., {Hugo}, B.,
  {Mirmont}, M., {Oonk}, J.~B.~R., {Prandoni}, I., {Rafferty}, D., {Sabater},
  J., {Smirnov}, O., {van Weeren}, R.~J., {White}, G.~J., {Atemkeng}, M.,
  {Bester}, L., {Bonnassieux}, E., {Br{\"u}ggen}, M., {Brunetti}, G.,
  {Chy{\.z}y}, K.~T., {Cochrane}, R., {Conway}, J.~E., {Croston}, J.~H.,
  {Danezi}, A., {Duncan}, K., {Haverkorn}, M., {Heald}, G.~H., {Iacobelli}, M.,
  {Intema}, H.~T., {Jackson}, N., {Jamrozy}, M., {Jarvis}, M.~J., {Lakhoo}, R.,
  {Mevius}, M., {Miley}, G.~K., {Morabito}, L., {Morganti}, R., {Nisbet}, D.,
  {Orr{\'u}}, E., {Perkins}, S., {Pizzo}, R.~F., {Schrijvers}, C., {Smith},
  D.~J.~B., {Vermeulen}, R., {Wise}, M.~W., {Alegre}, L., {Bacon}, D.~J., {van
  Bemmel}, I.~M., {Beswick}, R.~J., {Bonafede}, A., {Botteon}, A., {Bourke},
  S., {Brienza}, M., {Calistro Rivera}, G., {Cassano}, R., {Clarke}, A.~O.,
  {Conselice}, C.~J., {Dettmar}, R.~J., {Drabent}, A., {Dumba}, C., {Emig},
  K.~L., {En{\ss}lin}, T.~A., {Ferrari}, C., {Garrett}, M.~A.,
  {G{\'e}nova-Santos}, R.~T., {Goyal}, A., {G{\"u}rkan}, G., {Hale}, C.,
  {Harwood}, J.~J., {Heesen}, V., {Hoeft}, M., {Horellou}, C., {Jackson}, C.,
  {Kokotanekov}, G., {Kondapally}, R., {Kunert-Bajraszewska}, M., {Mahatma},
  V., {Mahony}, E.~K., {Mandal}, S., {McKean}, J.~P., {Merloni}, A., {Mingo},
  B., {Miskolczi}, A., {Mooney}, S., {Nikiel-Wroczy{\'n}ski}, B., {O'Sullivan},
  S.~P., {Quinn}, J., {Reich}, W., {Roskowi{\'n}ski}, C., {Rowlinson}, A.,
  {Savini}, F., {Saxena}, A., {Schwarz}, D.~J., {Shulevski}, A., {Sridhar},
  S.~S., {Stacey}, H.~R., {Urquhart}, S., {van der Wiel}, M.~H.~D., {Varenius},
  E., {Webster}, B., and {Wilber}, A. (2019).
\newblock {The LOFAR Two-metre Sky Survey. II. First data release}.
\newblock {\em \aap}, 622:A1.

\bibitem[{Thompson} et~al., 2017]{tms17}
{Thompson}, A.~R., {Moran}, J.~M., and {Swenson}, George~W., J. (2017).
\newblock {\em {Interferometry and Synthesis in Radio Astronomy, 3rd Edition}}.

\bibitem[{van Weeren} et~al., 2019]{wee19}
{van Weeren}, R.~J., {de Gasperin}, F., {Akamatsu}, H., {Br{\"u}ggen}, M.,
  {Feretti}, L., {Kang}, H., {Stroe}, A., and {Zandanel}, F. (2019).
\newblock {Diffuse Radio Emission from Galaxy Clusters}.
\newblock {\em \ssr}, 215(1):16.

\bibitem[{Venturi} et~al., 2007]{ven07}
{Venturi}, T., {Giacintucci}, S., {Brunetti}, G., {Cassano}, R., {Bardelli},
  S., {Dallacasa}, D., and {Setti}, G. (2007).
\newblock {GMRT radio halo survey in galaxy clusters at z = 0.2-0.4. I. The
  REFLEX sub-sample}.
\newblock {\em A\&A}, 463:937--947.

\bibitem[{Venturi} et~al., 2008]{ven08}
{Venturi}, T., {Giacintucci}, S., {Dallacasa}, D., {Cassano}, R., {Brunetti},
  G., {Bardelli}, S., and {Setti}, G. (2008).
\newblock {GMRT radio halo survey in galaxy clusters at z = 0.2-0.4. II. The
  eBCS clusters and analysis of the complete sample}.
\newblock {\em A\&A}, 484:327--340.

\end{thebibliography}

%%%%%%%%%%%%%%%%%%%%%%%%%%%%%%%%%%%%%%%%%%%%%%%%%%

\end{document}